\documentclass[aip,reprint,twocolumn,groupedaddress,superscriptaddress,showpacs]{revtex4-1}

\usepackage{amsmath}
\usepackage[hidelinks]{hyperref}
\usepackage[all]{hypcap}
\usepackage{graphicx}
\usepackage[caption=false]{subfig}
\usepackage{siunitx}
\usepackage{epstopdf}

\hypersetup{
	colorlinks   = true, 
	urlcolor     = blue, 
	linkcolor    = blue, 
	citecolor   = blue 
}

\makeatletter
\newcommand{\mypm}{\mathbin{\mathpalette\@mypm\relax}}
\newcommand{\@mypm}[2]{\ooalign{%
		\raisebox{.1\height}{$#1+$}\cr
		\smash{\raisebox{-.6\height}{$#1-$}}\cr}}
\makeatother

\begin{document}

\title{Upscaling High-Quality CVD Graphene Devices to 100 Micron-Scale and Beyond
}

\author{Timothy J. Lyon}
 \email{tlyon@wisc.edu}
 \altaffiliation[Also at ]{Department of Physics, University of Wisconsin-Madison.}
\author{Jonas Sichau}
\author{August Dorn}
\affiliation{
Center for Hybrid Nanostructures (CHyN), Department of Physics, University of Hamburg,
Jungiusstrasse 9-11, 20355 Hamburg, Germany\\
}

\author{Amaia Zurutuza}
\author{Amaia Pesquera}
\author{Alba Centeno}
 \homepage{http://www.graphenea.com}
\affiliation{Graphenea, Avenida de Tolosa 76, 20018 Donostia-San Sebastian, Spain\\
}

\author{Robert H. Blick}
 \homepage{http://www.nanomachines.com}
\affiliation{
Center for Hybrid Nanostructures (CHyN), Department of Physics, University of Hamburg,
Jungiusstrasse 9-11, 20355 Hamburg, Germany\\
}

\date{\today}

\begin{abstract}
  \noindent 
We describe a method for transferring ultra large-scale CVD-grown graphene sheets. These samples can be fabricated as large as several \SI{}{\centi\meter\squared} and are characterized by magneto-transport measurements on SiO$_2$ substrates. The process we have developed is highly effective and limits damage to the graphene all the way through metal liftoff, as shown in carrier mobility measurements and the observation of the quantum Hall effect. The charge-neutral point is shown to move drastically to near-zero gate voltage after a 2-step post-fabrication annealing process, which also allows for greatly diminished hysteresis.
\end{abstract}

\pacs{72.80.Vp, 81.05.ue.}
\maketitle

Graphene is well-known for its desirable electrical properties,~\cite{neto2009electronic} and with all of the intense focus on research in the material, new applications are being constantly discovered. While graphene is known to have its best electrical properties when it is single-crystalline and suspended,~\cite{bolotin2008ultrahigh,du2008approaching} those attributes are currently not feasible for mass-produced devices. There have been very exciting developments related to improving non-suspended graphene's mobility on more exotic substrates such as hBN,~\cite{dean2010boron,banszerus2015ultrahigh} as well as improving the chemical vapor deposition (CVD) growth process for producing better quality devices with large grain sizes.~\cite{petrone2012chemical,yan2012toward,banszerus2015ultrahigh} However, these developments are currently difficult to scale up and automate. 

On the other hand, CVD-grown graphene is still the best way to repeatably produce large areas of monolayer graphene, and SiO$_2$ is a well-known substrate that is already integrated into many processes from semiconductor physics to MEMS,~\cite{bunch2007electromechanical,zang2015graphene} and beyond.
Previous work has shown how to transfer large areas of CVD-grown graphene onto arbitrary substrates~\cite{liang2011toward,suk2011transfer,song2013general,pimenta2014direct} and remove
contaminants.~\cite{liang2011toward,her2013graphene} However, previous CVD-grown graphene on SiO$_2$ devices do not combine the desirable properties of high enough quality electrical characteristics to display the quantum Hall effect (QHE), a charge neutral point (CNP) near zero gate voltage, and a large device size, with typical finished devices being on the order of \SI{10}{\micro\meter}.~\cite{sun2014growth} Large-scale integration of easily manufactured, high-quality graphene devices is desirable in many different applications, such as graphene transistors, broad-band optical modulators~\cite{liu2011graphene} and \si{\tera\hertz} antennas.~\cite{tantiwanichapan2013graphene}

Challenges arise when fabricating high-quality CVD-grown graphene devices, primarily due to contaminants of all kinds easily attaching to graphene. With each step, much care must be taken to remove any existing organic or inorganic contaminants on the graphene as well as prevent new contaminants from being attached. This paper presents a fabrication method that produces devices with CVD-grown graphene on SiO$_2$ that are hundreds of microns in size and display the QHE.

A sheet of copper is supplied with CVD-grown graphene on only one side. Other sources of CVD graphene on copper typically have graphene on both sides of the foil, and in that case an oxygen plasma can be used to remove graphene from one side. A wet transfer process is used to place the graphene on the substrate, with cleaning steps to remove inorganic and organic contaminants from the bottom of the graphene before transfer. These steps are similar to the method detailed by Liang \textit{et al.}, which itself includes a ``modified RCA clean'' process.~\cite{liang2011toward}

It is important to be sure that all contaminants from the graphene are removed, as any resist residues can act as dopants and scattering centers to degrade electrical performance~\cite{bolotin2008ultrahigh,du2008approaching} as well as increase contact resistance.~\cite{li2013ultraviolet,lee2013clean} Acetone or other common solvents are not sufficient to remove residues from resists such as poly(methyl methacrylate) (PMMA).
A common technique to clean small graphene devices is to use current annealing,~\cite{moser2007current} especially when the graphene is suspended, as it can effectively evaporate many contaminants and allow the graphene to self-heal.~\cite{barreiro2013understanding} Other cleaning methods include mechanical cleaning with atomic force microscopy,~\cite{goossens2012mechanical} exposure to ultra-violet light,~\cite{chen2012uv} ozone treatment,~\cite{chen2012uv,li2013ultraviolet} and annealing in vacuum~\cite{pirkle2011effect} or with gas flow.~\cite{ishigami2007atomic} In the process detailed in this paper, the graphene is cleaned before transfer to a clean substrate, and then cleaned again after fabrication by a 2-step thermal annealing process.

\begin{figure}[htbp]
	\includegraphics[width=86mm]{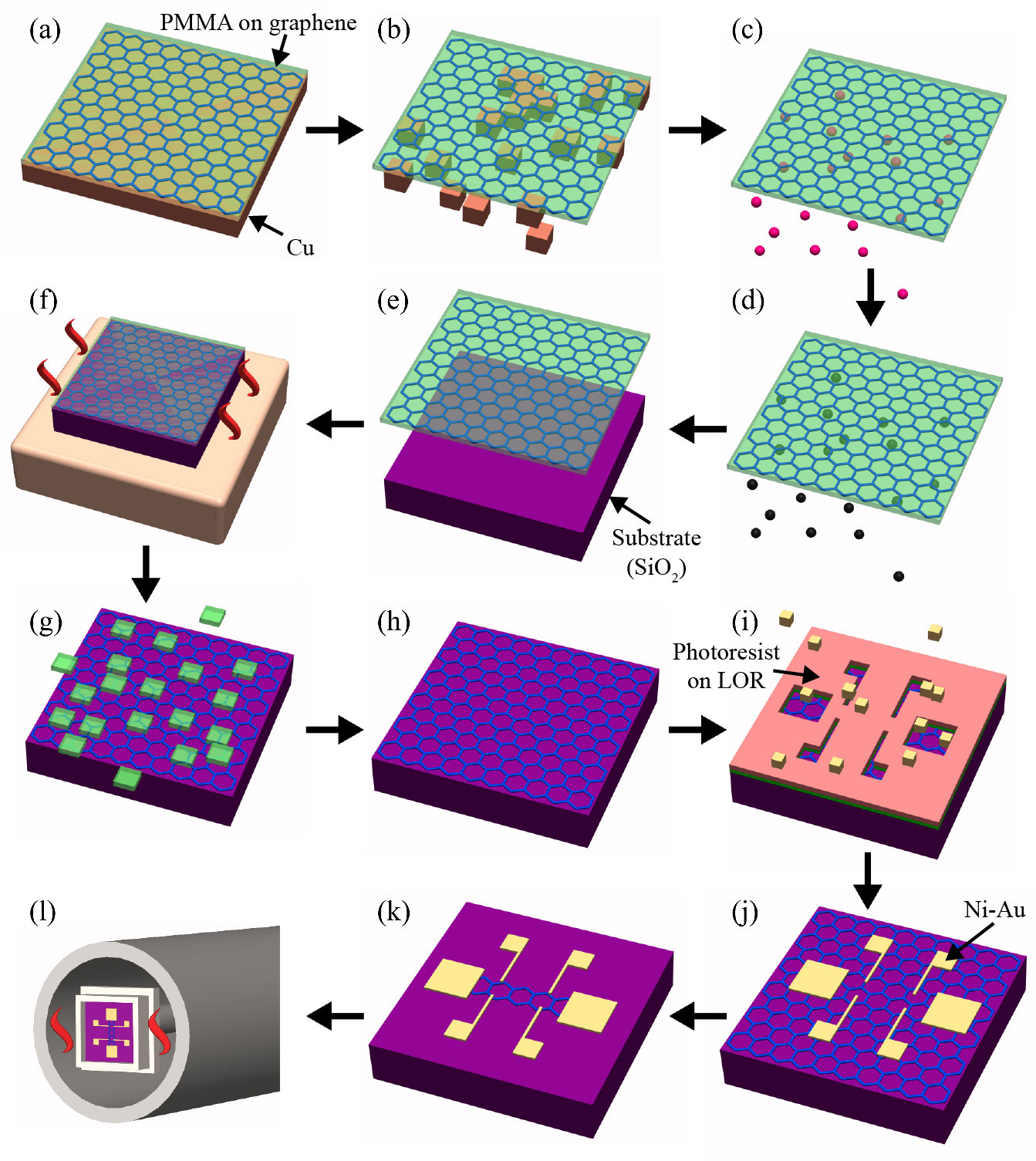}
	\caption{ \label{fig:3D_final}\footnotesize Flow diagram of the sample preparation. PMMA is spun onto the graphene copper stack (a), before the copper is etched away (b). The graphene is cleaned in multiple steps removing inorganic (c) and organic (d) contaminants, scooped up with a SiO$_2$ wafer piece (e) and then dried (f). The PMMA layer is removed with acetic acid (g) and the result is shown in (h). For the definition of the Hall contacts, the sample is coated with layers of LOR and photoresist, exposed and devoloped, before Ni and Au are deposited as adhesion and contact materials, respectively (i). After lift-off (j), the excess graphene is etched away (k) and the sample is mounted in a probe and baked under vacuum (l).}
\end{figure}

The size of the graphene/copper foil pieces used here is $ \SI{1}{\centi\meter} \times \SI{1}{\centi\meter}$, however, the process can easily be upscaled to sizes of several \si{\centi\meter^2}. PMMA with a molecular weight of 950K is spun onto the graphene/copper foil and dried by air (Fig.~\hyperref[fig:3D_final]{\ref{fig:3D_final}(a)}). A solution of Fe(NO$_3$)$_3$ is prepared with \SI{5}{g}/\SI{100}{ml} of DI water and the PMMA/graphene/copper stack is placed to float in the solution with the copper side down, then left for at least 10 hours for the copper to etch away (Fig.~\hyperref[fig:3D_final]{\ref{fig:3D_final}(b)}). Subsequently, the graphene stack is transferred to clean DI water at least twice, waiting 5 minutes after each transfer. Similarly to Liang \textit{et al.}, the RCA clean step referred to as Standard Clean 2 is next, with a HCl/H$_2$O$_2$/H$_2$O solution prepared in a 1:1:20 ratio. Once again, the stack is placed on this solution to clean inorganic contaminants not removed by the Fe(NO$_3$)$_3$, such as oxides, for 15 minutes (Fig.~\hyperref[fig:3D_final]{\ref{fig:3D_final}(c)}). The stack is then once again transferred to clean beakers of DI water at least twice for 5 minutes each. Finally, the RCA clean step known as Standard Clean 1 (SC-1) is performed with a NH$_4$OH/H$_2$O$_2$/H$_2$O solution in a 1:1:100 ratio for 5 minutes (Fig.~\hyperref[fig:3D_final]{\ref{fig:3D_final}(d)}) to remove organic contaminants. The cleaning time and chemical concentrations of H$_2$O$_2$ and NH$_4$OH in SC-1 are lower than reported in the modified RCA clean process in order to avoid bubble formation. Bubbles that get under the graphene stack can lead to the graphene tearing and are very difficult to remove before transfer. After the SC-1 step is complete, the graphene is again transferred to clean beakers of DI-water twice, waiting 5 minutes after each transfer.

\begin{figure}
	\includegraphics[width=86mm]{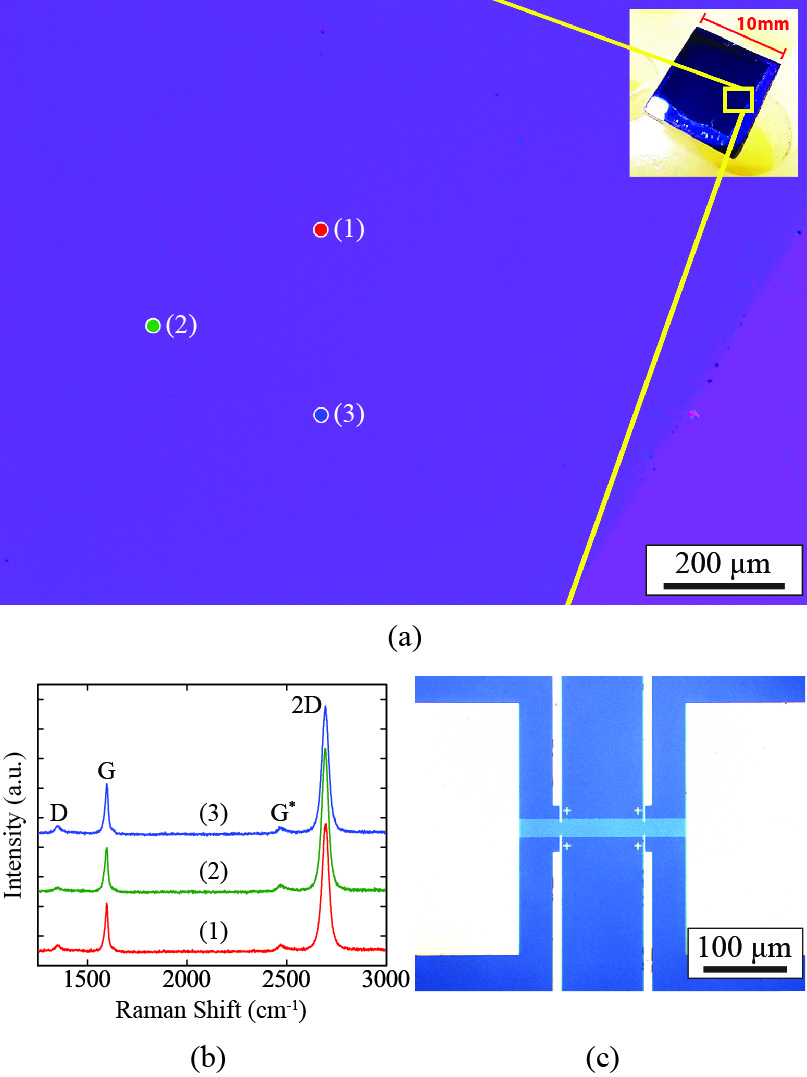}
	\caption{\label{fig:fig2} \footnotesize (a) Optical microscopy image of a large clean area of graphene (dark purple) on a SiO$_2$ substrate (light purple). Inset: Picture of a $ \SI{1}{\centi\meter} \times \SI{1}{\centi\meter} $ SiO$_2$ wafer with CVD graphene on top (dark blue). (b) Raman spectra with a \SI{532}{\nano\meter} laser taken at the colored and numbered spots shown in (a), displayed with an artificial offset for clarity. The defect peak D is very small in all three measurements, which speaks for a high quality of graphene, and the G$^{*}$ peak is in the expected location. Also, due to the relative height of the G and 2D peaks, which is about one third, we can confirm that we see a monolayer of graphene.~\cite{ferrari2006raman} (c) Optical microscopy image of a \SI{200}{\micro\meter} long and \SI{22}{\micro\meter} wide graphene strip (light blue) on an SiO$_2$ substrate (dark blue) with a contact layout typical for Hall measurements. (Graphene contrast increased for visual effect in (a) and (c)).}
\end{figure}

Wafers of thermally grown SiO$_2$ on $p$-doped Si are prepared by carefully cutting into appropriately sized pieces with a diamond scribe. Any contaminants are cleaned off with sonication in acetone and then isopropanol before the wafers are dried with a N$_2$ gun. The SiO$_2$ is then treated with O$_2$ plasma in order to make the surface more hydrophilic~\cite{nagashio2011electrical}, so the graphene will stick to it more readily and minimize breakage. Within one minute of the substrate's O$_2$ plasma exposure, the PMMA/graphene stack is scooped up and it is all dried in an oven set to \SI{150}{\celsius} for 15 minutes (Fig.~\hyperref[fig:3D_final]{\ref{fig:3D_final}(e)} and \hyperref[fig:3D_final]{(f)}).
Finally, the PMMA is removed with acetic acid (Fig.~\hyperref[fig:3D_final]{\ref{fig:3D_final}(g)}), which more cleanly removes PMMA residue than acetone, while at the same time not attacking either the graphene or the SiO$_2$ substrate.~\cite{her2013graphene}
At this point, the graphene is cleanly transferred to the substrate with minimal cracks or tears (Fig.~\hyperref[fig:3D_final]{\ref{fig:3D_final}(h)}), with most defects existing previous to the transfer process. Many other defects can be explained by bumps, folds, or other surface features in the copper foil that the graphene was grown on. These features make it much less likely for the graphene to transfer tear-free, making it very important to keep the copper foil as flat as possible. As shown in Fig.~\hyperref[fig:fig2]{\ref{fig:fig2}(a)} and~\hyperref[fig:fig2]{\ref{fig:fig2}(b)}, Raman spectra at \SI{532}{\nano\meter} were taken in multiple locations to verify reproducibility of the measurement and uniformity of the graphene monolayer.

Once the graphene is on the target substrate, it must still have metal contacts deposited as well as be patterned into whatever shape is desired. Both of these steps must ideally be done without damaging or contaminating the graphene any more than absolutely necessary. The contacts are defined using photolithography, with a layer of Microchem LOR 5A used between the photoresist and graphene. This bi-layer stack tends to remove much more cleanly than photoresist alone~\cite{nath2014achieving} and after removal can result in high-quality devices. In our observation, it also results in a higher final yield of usable devices due to less undesired graphene removal during liftoff. Without a protective layer of LOR, negative photoresists tended to result in much worse yield than positive resists, possibly due to crosslinking of polymer chains during exposure. After exposure and development, a \SI{5}{\nano\meter} layer of Ni is deposited by physical vapor deposition as an adhesion layer, followed by \SI{50}{\nano\meter} of Au (Fig.~\hyperref[fig:3D_final]{\ref{fig:3D_final}(i)}). Liftoff is then performed in a two-phase process. First, the sample is soaked in a bath of acetone, which removes the photoresist and lifts off the excess metal. Since an ultrasonic bath will damage the graphene, excess metal is instead gently removed by squeezing a pipette to agitate the acetone. Second, after all excess metal is removed, the LOR layer is removed by being soaked in Microposit Remover 1165 for five minutes. Finally, the sample is rinsed with isopropanol and dried with a N$_2$ gun. The result is shown in Fig.~\hyperref[fig:3D_final]{\ref{fig:3D_final}(j)}.

For this sample, excess graphene is removed by using photolithography to cover the areas of graphene to be protected from O$_2$ plasma. The sample is then chemically cleaned for the final time with acetone and isopropanol, then dried with N$_2$. Fig.~\hyperref[fig:3D_final]{\ref{fig:3D_final}(k)} and~\hyperref[fig:fig2]{\ref{fig:fig2}(c)} show the final result.

\begin{figure}
	\includegraphics[width=86mm]{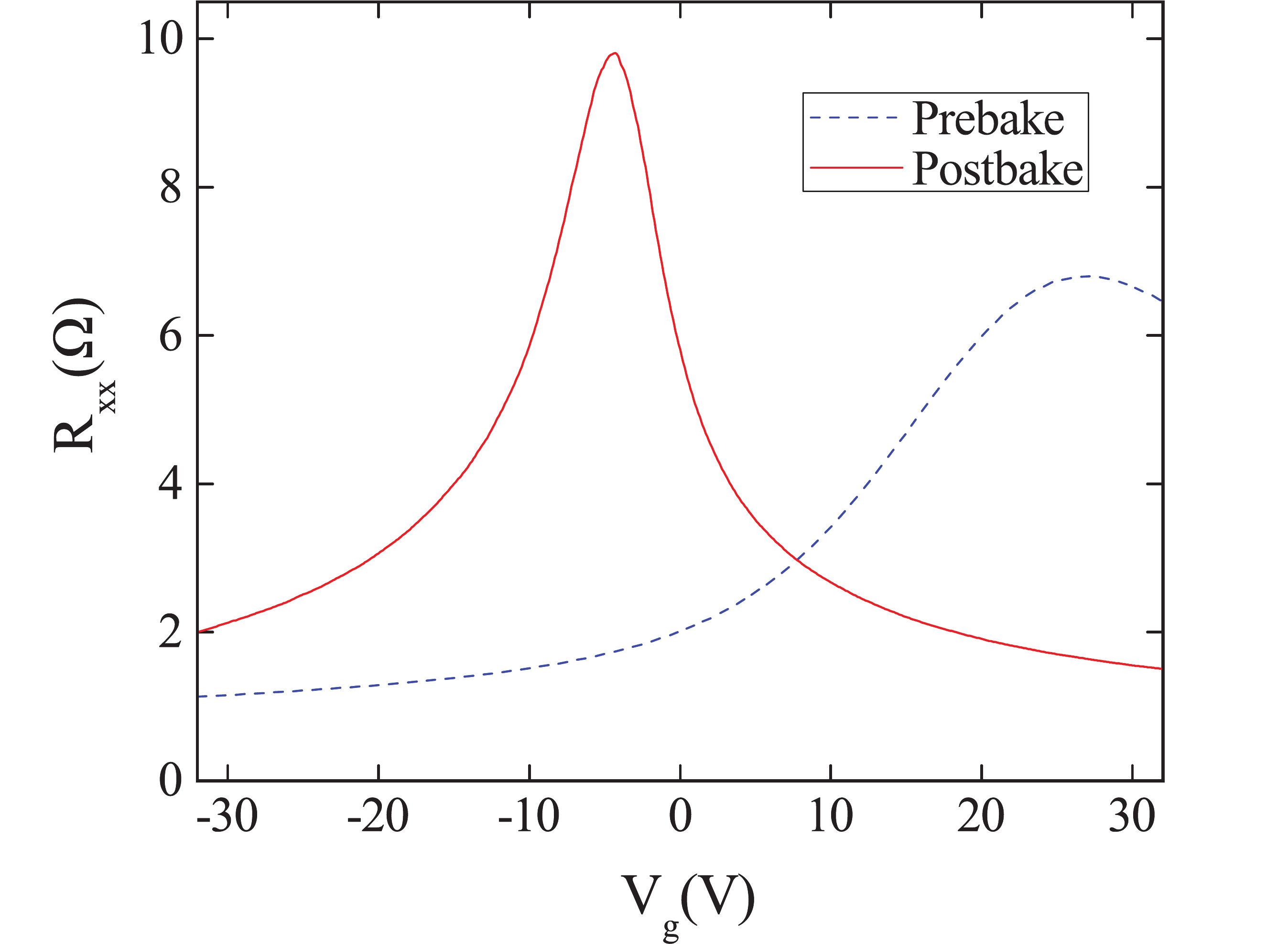}
	\caption{\label{fig:fig3_baking} \footnotesize Longitudinal resistance $ R_{\textrm{xx}} $ of the graphene strip versus gate voltage $ V_\textrm{g} $ before (dashed blue line) and after (solid red line) baking out the sample. The measurements were recorded at $ T = \SI{4.2}{\kelvin} $. A possible explanation for the postbake shift of the CNP of over \SI{30}{\volt} is the removal of contaminants on the graphene that act as dopants, originating from resist residues or the atmosphere.}
\end{figure}

The sample is then mounted in a probe with air pumped out and replaced with a small quantity of He as an exchange gas. The longitudinal resistance is measured over a range of applied gate voltages at a temperature of \SI{4.2}{\kelvin} at zero magnetic field. The sample is annealed under vacuum for 16 hours at \SI{350}{\celsius} to remove all water, including water trapped between the graphene and substrate. The sample is then removed from vacuum and quickly mounted in a probe, and once again annealed in a tube oven under vacuum (Fig.~\hyperref[fig:3D_final]{\ref{fig:3D_final}(l)}) at \SI{140}{\celsius} for 72 hours to remove any water absorbed from the atmosphere during mounting. During this time, the 2-point resistance of the Hall bar is observed to steadily increase from approximately \SI{25}{\kilo\ohm} to a new maximum of \SI{57}{\kilo\ohm}. Fig.~\ref{fig:fig3_baking} shows the increase in peak longitudinal resistance along with a dramatic shift of the charge neutral point (CNP) from $V_{\textrm{g}}$=\SI{27}{\volt} to \SI{-4}{\volt} before and after annealing, respectively, which is evidential of the concentration of charged impurities being significantly reduced.~\cite{adam2007self} The large peak shift is indicative of the removal of a majority of $p$-type dopants such as water~\cite{novoselov2004electric} and the settling of the CNP at a negative gate voltage may be due to doping by the metal contacts or contact doping by SiO$_{2}$ and any contaminants trapped between the graphene and metal during processing.~\cite{giovannetti2008doping,nagashio2011density} When the sample is re-exposed to atmosphere, the CNP drifts toward the prebake value, though performing the baking process once more moves the CNP back to near-zero gate voltage. Thus, a capping layer is necessary to prevent adsorbates from altering the electrical characteristics of the graphene when exposed to air.

\begin{figure}
	\includegraphics[width=86mm]{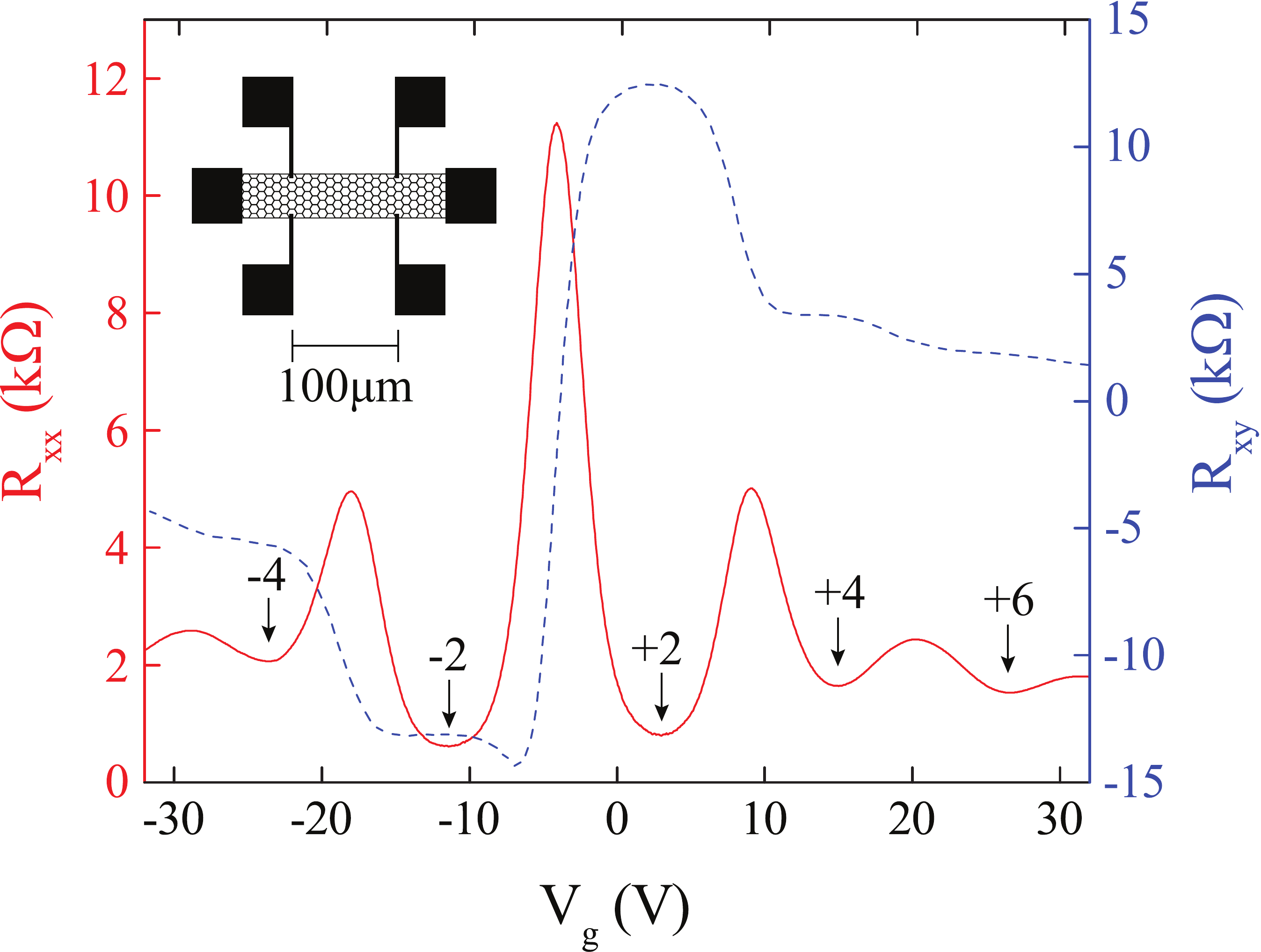}
	\caption{\label{fig:fig4_qhall} \footnotesize The longitudinal (solid red line) and Hall resistance (dashed blue line) versus the gate voltage $V_{\textrm{g}}$ at fixed magnetic field $B=\SI{8}{\tesla}$ measured at \SI{4.2}{\kelvin}, obtained through a standard 4-point-measurement approach, show multiple quantum Hall levels with filling factors labeled. Inset: Schematic of the graphene strip and contacts, with a distance between the longitudinal contacts of \SI{100}{\micro\meter}. }
\end{figure}

The CNP changes by at most \SI{0.1}{\volt}, depending on the direction the gate voltage is swept. When there is still a significant amount of water on the substrate, this hysteresis can easily be on the order of a few or even tens of volts.~\cite{nagashio2011electrical} This effect is typical of graphene on SiO$_2$ and the relatively low hysteresis seen in this sample implies that most of the water has been removed by the previous annealing step.~\cite{lafkioti2010graphene}

\begin{figure}
	\includegraphics[width=86mm]{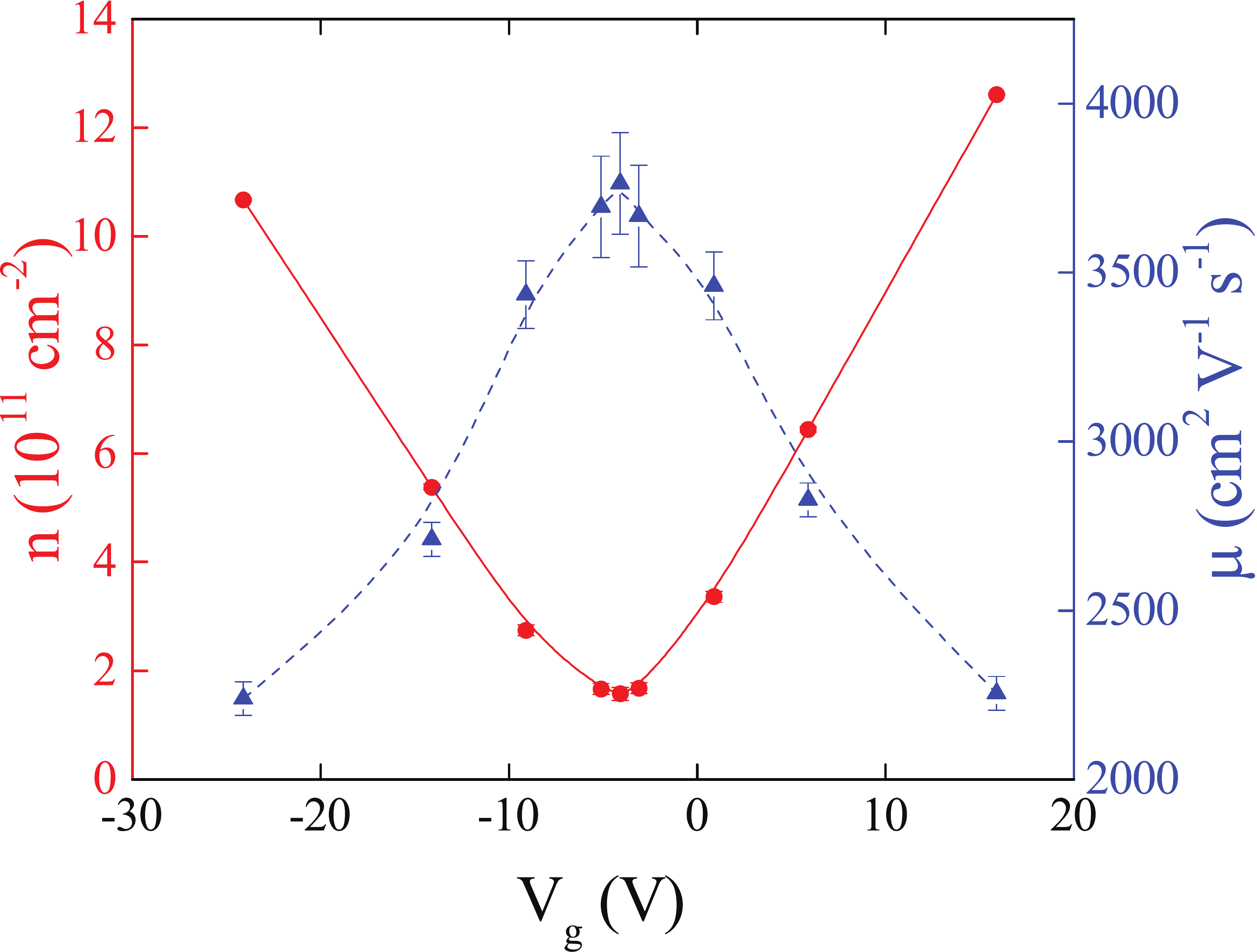}
	\caption{\label{fig:fig5_n_u} \footnotesize The charge carrier density (solid red line) and mobility (dashed blue line) versus the gate voltage $V_{\textrm{g}}$ at \SI{4.2}{\kelvin} after the final annealing step. The values were obtained by performing magnetoresistance measurements at the displayed values for $V_{\textrm{g}}$. Points are connected by B-spline curves. The charge carrier density is linear with respect to gate voltage away from the CNP, which implies the gate oxide provides the dominant capacitive effect which is expected due to the oxide thickness.~\cite{fang2007carrier} The rounding of the charge carrier density graph near the CNP is due to charged electron/hole puddles induced by charged impurities, including those in the substrate.~\cite{adam2007self}}
\end{figure}

The magnetic field is then set to \SI{8}{\tesla} and the gate voltage is scanned once more at \SI{4.2}{\kelvin}. Multiple quantum Hall levels for monolayer graphene were seen, as shown in Fig.~\ref{fig:fig4_qhall}. Clear plateaux for the Hall resistance, $R_{\textrm{xy}}$, at filling factors of $\nu=\mypm 4(n+1/2)$ and Landau level index $n=0,1,...$ are seen with corresponding drops in the longitudinal resistance, $R_{\textrm{xx}}$.~\cite{zhang2005experimental} Further magnetoresistance measurements were performed to find charge carrier density and mobility at nine different gate voltages with the result shown in Fig.~\ref{fig:fig5_n_u}. At the CNP, the mobility is measured to be around $ \mu_{\textrm{CNP}} = \SI{3760}{\centi\meter\squared\per(\volt\second)} $ and the charge carrier density is $ n_{\textrm{CNP}} = \SI{1.58e11}{\centi\meter^{-2}} $. These measurements show that the fabrication method presented here results in a mobility at the charge neutral point that approaches values obtained when using exfoliated graphene on SiO$_{2}$,~\cite{huang2015reliable} implying that the graphene is very clean and has been minimally damaged by the sample production process.

In summary, we have presented a procedure that results in clean large-area graphene devices of high quality on a SiO$_2$ substrate. The 2-step liftoff aids greatly in unbroken large graphene structures, and the 2-step annealing process as well as other cleaning steps result in a realization of near-ideal mobilities and a low hysteresis with observation of quantum Hall levels. Hence, our processing approach now enables large-scale integration of high-quality graphene layers in devices such as THz-emitters, thermo-power couplers, and possibly flexible thin-film sensors.

The authors acknowledge funding support by the MURI’08 of the Air Force Office of Scientific Research (AFOSR, Award No. FA9550-08-1-0337), and the Center for Ultrafast Imaging (CUI) at the University of Hamburg, which is sponsored by the Deutsche Forschungsgemeinschaft (DFG) through grant EXC-1074.

\bibliography{fabrication}

\end{document}